\documentclass[aps,prd,superscriptaddress,showpacs,preprint,amsmath,amssymb]{revtex4}
\usepackage{graphicx, bm}
\usepackage[usenames]{color}

\begin{document}

\draft
\title{Model-independent study for the tau-neutrino electromagnetic dipole moments in $e^+e^- \to \nu_\tau \bar \nu_\tau \gamma$ at the CLIC}

\author{ A. Guti\'errez-Rodr\'{\i}guez\footnote{alexgu@fisica.uaz.edu.mx}}
\affiliation{\small Facultad de F\'{\i}sica, Universidad Aut\'onoma de Zacatecas\\
         Apartado Postal C-580, 98060 Zacatecas, M\'exico.\\}

\author{M. K\"{o}ksal\footnote{mkoksal@cumhuriyet.edu.tr}}
\affiliation{\small Deparment of Optical Engineering, Cumhuriyet University, 58140, Sivas, Turkey.\\}

\author{A. A. Billur\footnote{abillur@cumhuriyet.edu.tr}}
\affiliation{\small Deparment of Physics, Cumhuriyet University, 58140, Sivas, Turkey.\\}

\author{ M. A. Hern\'andez-Ru\'{\i}z\footnote{mahernan@uaz.edu.mx}}
\affiliation{\small Unidad Acad\'emica de Ciencias Qu\'{\i}micas, Universidad Aut\'onoma de Zacatecas\\
         Apartado Postal C-585, 98060 Zacatecas, M\'exico.\\}

\date{\today}

\begin{abstract}

We conduce a study to probe the sensitivity of the process $e^+e^-\rightarrow (\gamma, Z) \to \nu_\tau \bar \nu_\tau \gamma$
to the total cross section, the magnetic moment and the electric dipole moment of the tau-neutrino in a model-independent way.
For this study, the beam polarization facility at the Compact Linear Collider (CLIC) along with the typical center-of-mass
energies $\sqrt{s}=380-3000\hspace{0.8mm}GeV$ and integrated luminosities ${\cal L}=10-3000\hspace{0.8mm}fb^{-1}$ is considered.
We estimate the sensitivity at the $95\%\hspace{1mm}$ Confidence Level (C.L.) and systematic uncertainties $\delta_{sys}=0, 5,
10\hspace{1mm}\%$ on the dipole moments of the tau-neutrino. It is shown that the process under consideration $e^+e^-\rightarrow
(\gamma, Z) \to \nu_\tau \bar \nu_\tau \gamma$ is a good prospect for study the dipole moments of the tau-neutrino at the CLIC.
Also, our study illustrates the complementarity between CLIC and other $e^+e^-$ and $pp$ colliders in probing extensions
of the Standard Model, and shows that the CLIC at high energy and high luminosity provides a powerful means to sensitivity estimates
for the electromagnetic dipole moments of the tau-neutrino.

\end{abstract}


\vspace{5mm}

\maketitle

\section{Introduction}

Investigations of the theory and phenomenology of neutrino electromagnetic properties continue to be a very active
field of interests to both theoretical and experimental physicists. In particular, the study of the Magnetic Moment (MM)
and the Electric Dipole Moment (EDM) of the neutrino has been challenging High Energy Physics community, both Theoretical
and Experimental in recent decades. In the original formulation of the Standard
Model (SM) \cite{Glashow,Weinberg,Salam} neutrinos are massless particles with zero MM. However, neutrino flavour oscillation
experiments from several sources indicate that neutrinos have non-zero mass, which indicates the necessity of extending
the SM to accommodate massive neutrinos. In the minimal extension of the SM to incorporate the neutrino mass, the MM of
the neutrino is known to be developed in one loop calculation \cite{Fujikawa,Shrock}, and the non-zero mass of
the neutrino is essential to get a non-vanishing MM. Furthermore, the SM predicts CP violation, which is necessary
for the existence of the EDM of a variety physical systems. The EDM provides a direct experimental
probe of CP violation \cite{Christenson,Abe,Aaij}, a feature of the SM and beyond the SM (BSM) physics. The signs of new physics
can be analyzed by investigating the electromagnetic dipole moments of the tau-neutrino, such as its MM and EDM.

The present best upper limits on the MM and the EDM of the neutrinos, either set directly by experiments or inferred indirectly
from observational evidences combined with theoretical arguments, are several orders of magnitude larger than the predictions
of the minimal extension of the SM \cite{Fujikawa,Shrock,Data2016}. Therefore, if any direct experimental confirmation of non-zero
MM is obtained in the laboratory experiments, it will open a window to “new physics. In addition, the dipole moments with the copious
amount of neutrinos in the Universe will have significant implications for astrophysics and cosmology, as well as terrestrial neutrino
experiments \cite{Cisneros,Raffelt}. One of the most sensitive experimental observables to the CP violation BSM is the EDM \cite{Yamanaka,Yamanaka1,Chupp,Engel}. The search for new sources of CP violation BSM is currently one of the most important
fundamental problems of particle physics to be solved. A. Sakharov proposed a solution to this problem \cite{Sakharov}, the 
present interaction has to violate a fundamental symmetry of nature: the CP symmetry. The excess of matter over antimatter, 
or the baryon number asymmetry, was generated in the early Universe by a theory satisfy Sakharov criteria.

Another interesting topics in neutrino physics is to determine its Dirac or Majorana nature. For respond to this, experimentalist are
exploring different reactions where the Majorana nature may manifest \cite{Zralek}. About this topic, the study of neutrino magnetic
moments is, in principle, a way to distinguish between Dirac and Majorana neutrinos since the Majorana neutrinos can only have flavor
changing, transition magnetic moments while the Dirac neutrinos can only have flavor conserving one.

The anomalous MM of the neutrino are being searched in reactor (MUNU, TEXONO and GEMMA) \cite{MUNU,TEXONO,GEMMA},
accelerator (LSND) \cite{Auerbach,DONUT}, and solar (Super-Kamiokande and Borexino) \cite{Liu,Borexino} experiments.
The current best sensitivity limits on the MM obtained in laboratory measurements are

\begin{equation}
\mu^{exp}_{\nu_e}= 2.9 \times 10^{-11}\mu_B, \hspace{3mm} 90\% \hspace{2mm} C.L. \hspace{5mm} \mbox{[GEMMA]} \hspace{5mm} \mbox{\cite{GEMMA}},
\end{equation}

\begin{equation}
\mu^{exp}_{\nu_\mu}= 6.8 \times 10^{-10}\mu_B, \hspace{3mm} 90\% \hspace{2mm} C.L. \hspace{5mm} \mbox{[LSND]} \hspace{5mm} \mbox{\cite{Auerbach}}.
\end{equation}

These sensitivity limits exceed by many orders of magnitude the minimally extended SM prediction given by

\begin{equation}
\mu_\nu=\frac{3eG_F m_{\nu_i}}{(8\sqrt{2}\pi^2)}\simeq 3.1\times 10^{-19}(\frac{m_{\nu_i}}{1 \hspace{1mm} eV})\mu_B,
\end{equation}

\noindent where $\mu_B=\frac{e}{2m_e}$ is the Bohr magneton \cite{Fujikawa,Shrock}.

The best world sensitivity bounds for the electric dipole moments $d_{\nu_e, \nu_\mu}$ \cite{Aguila} are:

 \begin{equation}
d_{\nu_e, {\nu}_\mu} < 2\times 10^{-21} (e cm), \hspace{5mm} 95\%\hspace{0.8mm}C.L..\\
\end{equation}

For the $\tau$-neutrino, the bounds on their dipole moments are less restrictive, and therefore it is worth investigating
in deeper way their electromagnetic properties. The tau-neutrino correspond to the more massive third generation of neutrinos
and possibly possesses the largest mass and the largest MM and EDM. As a consequence, this leaves space for the study of new
physics BSM.

A summary of experimental and theoretical limits on the dipole moments of the tau-neutrino are given in Table I of Ref. \cite{Gutierrez13}.
See Refs. \cite{Billur,Gutierrez12, Gutierrez11,Gutierrez10,Gutierrez9,Gutierrez8,Data2016,Gutierrez7,Sahin,Sahin1,Gutierrez6,Aydin,Gutierrez5,Gutierrez4,Gutierrez3,Keiichi,
Aytekin,Gutierrez2,Gutierrez1,DELPHI,Escribano,Gould,Grotch} for another limits on the MM and the EDM of the $\tau$-neutrino in different
context.

A central goal of the physics programme of the future lepton colliders is to complement the Large Hadron Collider (LHC) results and
also search for clues in BSM. The lepton colliders are designed to study the properties of the new particles and the
interactions they might undergo according to the vast amount of theories. Furthermore, the lepton colliders compared to the LHC have
a cleaner background, and it is possible to extract the new physics signals from the background more easily. In this regard, there is
currently an ongoing effort for the project named the Compact Linear Collider (CLIC) \cite{Accomando,Dannheim,Abramowicz}. When is
constructed and enter into operation, the $\gamma e^-$ and $\gamma\gamma$ collision modes will be studied. The CLIC
will be a multi-TeV collider and will be operate in three energy stages, corresponding to center-of-mass energies $\sqrt{s}= 380, 1500, 3000\hspace{0.8mm}GeV$, and it is an ideal machine to study new physics BSM.

Motivated by the extensive physical program of the CLIC, we conduce a comprehensive study to probe the sensitivity of the process
$e^+e^-\rightarrow (\gamma, Z) \to \nu_\tau \bar \nu_\tau \gamma$ to the total cross section, the MM and the EDM of the tau-neutrino
in a model-independent way. For the study, the beam polarization facility at the CLIC along with the typical center-of-mass energies
$\sqrt{s}=380, 1500, 3000\hspace{0.8mm}GeV$ and integrated luminosities ${\cal L}=10, 50, 100, 300, 500, 1000, 1500, 2000, 3000\hspace{0.8mm}fb^{-1}$
are considered. In addition, we estimate the sensitivity at the $95\%\hspace{1mm}C. L.$ and systematic uncertainties $\delta_{sys}=0, 5, 10\hspace{1mm}\%$
on the dipole moments of the $\tau$-neutrino. It is shown that the process under consideration $e^+e^-\rightarrow (\gamma, Z) \to \nu_\tau \bar \nu_\tau \gamma$ is a good prospect for study the dipole moments of the tau-neutrino at the CLIC. Furthermore, our study illustrates the complementarity
between CLIC and other $e^+e^-$ and $pp$ colliders in probing extensions of the SM, and shows that the CLIC at high energy and high
luminosity provides a powerful means to sensitivity estimates for the electromagnetic dipole moments of the tau-neutrino.

The content of this paper is organized as follows: In Section II, we study the total cross section and the dipole moments of the tau-neutrino
through the channel $e^+e^-\rightarrow (\gamma, Z) \to \nu_\tau \bar \nu_\tau \gamma$. Finally we conclude in Section III.

\section{The total cross section of the process $e^+ e^- \to \nu_\tau\bar\nu_\tau\gamma$ and dipole moments}

\subsection{Electromagnetic vertex $\nu_\tau\bar\nu_\tau\gamma$}

Theoretically the electromagnetic properties of neutrinos best studied and well understood are the MM and the EDM.
Despite that the neutrino is a neutral particle, neutrinos can interact with a photon through loop (radiative) diagrams.
However, a convenient way of studying its electromagnetic properties on a model-independent way is through the effective
neutrino-photon interaction vertex which is described by four independent form factors. The most general expression for
the vertex of interaction $\nu_\tau\bar\nu_\tau\gamma$ is given in Refs. \cite{Nieves,Kayser,Kayser1}. For the study of
the MM and the EDM of the tau-neutrino we following a focusing as the performed in our previous works \cite{Billur,Gutierrez13,
Gutierrez12,Gutierrez11,Gutierrez10,Gutierrez9,Gutierrez8, Gutierrez7, Gutierrez6,Gutierrez5,Gutierrez4, Gutierrez3,Gutierrez2,
Gutierrez1} with

\begin{equation}
\Gamma^{\alpha}=eF_{1}(q^{2})\gamma^{\alpha}+\frac{ie}{2m_{\nu_\tau}}F_{2}(q^{2})\sigma^{\alpha
\mu}q_{\mu}+ \frac{e}{2m_{\nu_\tau}}F_3(q^2)\gamma_5\sigma^{\alpha\mu}q_\mu +eF_4(q^2)\gamma_5(\gamma^\alpha-\frac{q\llap{/}q^\alpha}{q^2}),
\end{equation}

\noindent where $e$ is the electric charge of the electron, $m_{\nu_\tau}$ is the mass of the tau-neutrino, $q^\mu$
is the photon momentum, and $F_{1, 2, 3, 4}(q^2)$ are the four electromagnetic form factors of the neutrino. In general
the $F_{1, 2, 3, 4}(q^2)$ are independent form factors, and they are not physical quantities, but in the limit $q^2 \to 0$
they are quantifiable and related to the static quantities corresponding to charge radius, MM, EDM and anapole moment (AM)
of the Dirac neutrinos, respectively \cite{Escribano,Vogel,Bernabeu1,Bernabeu2,Dvornikov,Giunti,Broggini}. In this paper we
study the anomalous MM $\mu_{\nu_\tau}$ and the EDMt $d_{\nu_\tau}$ of the tau-neutrino, which are defined in terms of the
$F_2(q^2=0)$ and $F_3(q^2=0)$ independent form factor as follows:

\begin{eqnarray}
\mu_{\nu_\tau}&=&\Bigl(\frac{m_e}{m_{\nu_\tau}}\Bigr)F_2(0)\mu_B,\\
d_{\nu_\tau}&=&\Bigl(\frac{e}{2m_{\nu_\tau}}\Bigr) F_3(0),
\end{eqnarray}

\noindent as we mentioned above. The form factors corresponding to charge radius and the anapole moment, are not considered in this paper.

\subsection{Total cross section of the process $e^+ e^- \to \nu_\tau\bar\nu_\tau\gamma$ beyond the SM with unpolarized electron-positron beam}

The corresponding Feynman diagrams for the signal $e^+ e^- \to (\gamma, Z) \to \nu_\tau\bar\nu_\tau\gamma$ are given in Fig. 1.
The total cross section of the process $e^+ e^- \to \nu_\tau\bar\nu_\tau\gamma$ with unpolarized electron-positron beam is computed
using the CALCHEP 3.6.30 \cite{Calhep} package, which can computate the Feynman diagrams, integrate over multiparticle phase
space and event simulation. Furthermore, in order to select the events we implementing the standard isolation cuts, compatibly
with the detector resolution expected at CLIC:

\begin{eqnarray}
\begin{array}{c}
\hspace{2mm} p^\nu_T > 150\hspace{0.8mm}GeV,\\
\hspace{2mm} |\eta^\gamma| < 2.37,\\
\hspace{2mm} p^\gamma_T > 150\hspace{0.8mm}GeV,
\end{array}
\end{eqnarray}

\noindent we apply these cuts to reduce the background and to optimize the signal sensitivity. In Eq. (8), $p^\nu_T$ is the transverse
momentum of the final state neutrinos, $\eta^\gamma$ is the pseudorapidity and $p^\gamma_T$ is the transverse momentum of the photon.
The outgoing particles are required to satisfy these isolation cuts.

Formally, the $e^+ e^- \to (\gamma, Z) \to \nu_\tau\bar\nu_\tau\gamma$  cross section can be split into two parts:

\begin{equation}
\sigma=\sigma_{BSM} + \sigma_0,
\end{equation}

\noindent where $\sigma_{BSM}$ is the contribution due to BSM physics, which, in our case it comes from the anomalous vertex
$\nu_\tau\bar\nu_\tau\gamma$, while $\sigma_0$ is the SM prediction. The analytical expression for the squared amplitudes are
quite lengthy so we do not present it here. Following the form of Eq. (9), we present numerical fit functions for the total
cross section with respect to center-of-mass energy, with unpolarized electron-positron beam and in terms of the independent
form factors $F_2 (F_3)$.\\


$\bullet$ For $\sqrt{s}=380\hspace{0.8mm} GeV$.

\begin{eqnarray}
\sigma(F_2)&=&\bigl[(2.68 \times 10^{11}) F^4_2 + (1.97 \times 10^{4} ) F^2_2 + 0.041\bigr](pb),   \nonumber \\
\sigma(F_3)&=&\bigl[(2.68 \times 10^{11}) F^4_3 + (1.97 \times 10^{4} ) F^2_3 + 0.041\bigr](pb).
\end{eqnarray}

$\bullet$ For $\sqrt{s}=1.5\hspace{0.8mm} TeV$.

\begin{eqnarray}
\sigma(F_2)&=&\bigl[(3.32 \times 10^{13}) F^4_2 + (5.13 \times 10^{5} ) F^2_2 + 0.012\bigr](pb),   \nonumber \\
\sigma(F_3)&=&\bigl[(3.32 \times 10^{13}) F^4_3 + (5.13 \times 10^{5} ) F^2_3 + 0.012\bigr](pb).
\end{eqnarray}

\newpage

$\bullet$ For $\sqrt{s}=3\hspace{0.8mm} TeV$.

\begin{eqnarray}
\sigma(F_2)&=&\bigl[(1.49 \times 10^{14}) F^4_2 + (9.70 \times 10^{5} ) F^2_2 + 0.003\bigr](pb),   \nonumber \\
\sigma(F_3)&=&\bigl[(1.49 \times 10^{14}) F^4_3 + (9.70 \times 10^{5} ) F^2_3 + 0.003\bigr](pb).
\end{eqnarray}

\noindent It is worth mentioning that in equations for the total cross section (10)-(12), the coefficients of $F_2 (F_3)$ given
the anomalous contribution, while the independent terms of $F_2 (F_3)$ correspond to the cross section at $F_2=F_3=0$ and
represents the SM total cross section magnitude.

\subsection{Sensitivty estimates on the $\mu_{\nu_\tau}$ and $d_{\nu_\tau}$  with unpolarized electron-positron beam}

Based on the formulas given by Eqs. (10)-(12), we make model-independent sensitivity estimates for the total cross section of the signal
$\sigma_{BSM}(e^+e^- \rightarrow \nu_\tau \bar\nu_\tau\gamma)=\sigma_{BSM}(\sqrt{s},\hspace{0.8mm} \mu_{\nu_\tau},\hspace{0.8mm}d_{\nu_\tau})$,
as well as for the anomalous MM $\mu_{\nu_\tau}$ and EDM $d_{\nu_\tau}$ of the $\tau$-neutrino at the CLIC.
To carry out this task, we consider the acceptance cuts given in Eq. (8) and we take into account the systematic uncertainties
$\delta_{sys}=0, 5, 10\hspace{1mm}\%$ for the collider. In addition, to sensitivity estimates on the parameters of the process
$e^+e^- \rightarrow \nu_\tau \bar\nu_\tau \gamma$, we use the $\chi^2$ function \cite{Koksal1,Koksal2,Gutierrez13,Koksal,Ozguven,Billur,Sahin1,Billur1,Billur2}

\begin{equation}
\chi^2=\Biggl(\frac{\sigma_{SM}-\sigma_{BSM}(\sqrt{s}, \mu_{\nu_\tau}, d_{\nu_\tau})}{\sigma_{SM}\sqrt{(\delta_{st})^2
+(\delta_{sys})^2}}\Biggr)^2,
\end{equation}

\noindent where $\sigma_{BSM}(\sqrt{s}, \mu_{\nu_\tau}, d_{\nu_\tau})$ is the total cross section including
contributions from the SM and new physics, $\delta_{st}=\frac{1}{\sqrt{N_{SM}}}$ is the statistical error
and $\delta_{sys}$ is the systematic error. The number of events is given by $N_{SM}={\cal L}_{int}\times \sigma_{SM}$,
where ${\cal L}_{int}$ is the integrated CLIC luminosity.

As stated in the Introduction, to carry out our study we considered the typical center-of-mass energies $\sqrt{s}=380, 1500, 3000\hspace{0.8mm}GeV$
and integrated luminosities ${\cal L}=10, 50, 100, 300, 500, 1000, 1500, 2000, 3000\hspace{0.8mm}fb^{-1}$ of the CLIC.

We report in Figs. 2 and 3 the sensitivity on the signal cross section $e^+e^- \rightarrow (\gamma, Z) \to \nu_\tau \bar\nu_\tau \gamma$
at the CLIC as a function of the form factors $F_2 (F_3)$ and for different center-of mass energies $\sqrt{s}=380, 1500, 3000 \hspace{0.8mm}GeV$.
Clearly the total cross section is dominant for $\sqrt{s}=3000\hspace{0.8mm}GeV$ and for large values of the form factors $F_2 (F3)$, and
decreases as $F_2 (F_3)$ tends to zero, recovering the value of the SM as it is shown in Eq. (12).

Sensitivity contours at the $95\% \hspace{1mm}C.L.$ in the $F_3-F_2$ plane for the signal $e^+e^-\rightarrow \nu_\tau \bar \nu_\tau \gamma$
with center-of-mass energies $\sqrt{s}=380, 1500, 3000\hspace{0.8mm}GeV$ and luminosities ${\cal L}=10, 100, 500, 1500, 3000 \hspace{0.8mm}fb^{-1}$
are given in Figs. 4-6. As highlighted in Fig. 6, the three most sensitive contours for $F_2$ and $F_3$ they are the corresponding ones
for high energy and high luminosity of $\sqrt{s}=3000\hspace{0.8mm}GeV$ and ${\cal L}=3000 \hspace{0.8mm}fb^{-1}$.

As a final result on our sensitivity analysis, we stress the sensitivity estimates on the $\mu_{\nu_\tau}$ and $d_{\nu_\tau}$
via the channel $e^+ e^- \to \nu_\tau \bar\nu_\tau\gamma$ for $\sqrt{s}=380, 1500, 3000\hspace{0.8mm}GeV$,
${\cal L}=10, 50, 100, 300, 500, 1000, 1500, 2000, 3000 \hspace{0.8mm}fb^{-1}$, $\delta_{sys}=0, 5, 10\%$ at $90\%\hspace{0.8mm}C. L.$
and $95\%\hspace{0.8mm}C. L.$. We show our results in Tables I-III, where the better sensitivity on the dipole moments of the $\tau$-neutrino
projected for the CLIC are for $\sqrt{s}=3000\hspace{0.8mm}GeV$ and ${\cal L} = 3000 \hspace{0.8mm}fb^{-1}$:
$|\mu_{\nu_\tau}(\mu_B)|=2.103\times 10^{-7}$ and $|d_{\nu_\tau}(e cm)|=4.076\times10^{-18}$ at $90\%\hspace{0.8mmC.L.}$.

\begin{table}[!ht]
\caption{Sensitivity estimates on the $\mu_{\nu_\tau}$ magnetic moment and $d_{\nu_\tau}$ electric dipole moment via the process
$e^+ e^- \to \nu_\tau \bar\nu_\tau \gamma$ for $\sqrt{s}=380\hspace{0.8mm}GeV$ and $P_{e^-}=P_{e^+}=0\%$.}
\begin{center}
\begin{tabular}{|c|cc|cc|cc|}
\hline\hline
\multicolumn{7}{|c|}{ $90\%$ C.L. \hspace{5mm} $\sqrt{s}=380\hspace{0.8mm}GeV$ }\\
\hline\hline
\multicolumn{7}{|c|}{\hspace{2.5cm}  $\delta_{sys}=0\%$  \hspace{2.5cm}  $\delta_{sys}=5\%$  \hspace{2.5cm}  $\delta_{sys}=10\%$ }\\
\hline

\cline{1-7}  ${\cal L}\hspace{0.8mm}(fb^{-1})$  & \hspace{2mm} $|\mu_{\nu_\tau}(\mu_B)|$   & $\hspace{2mm} |d_{\nu_\tau}(e cm)|$
     & \hspace{2mm} $|\mu_{\nu_\tau}(\mu_B)|$   & $\hspace{2mm} |d_{\nu_\tau}(e cm)|$      & \hspace{2mm} $|\mu_{\nu_\tau}(\mu_B)|$
     & $\hspace{2mm} |d_{\nu_\tau}(e cm)|$         \\
\hline
  10   &\hspace{0.8mm}  $7.953\times 10^{-6}$    & \hspace{0.8mm}   $1.541\times 10^{-16}$  &  \hspace{0.8mm} $8.914\times 10^{-6}$ & \hspace{0.8mm}
  $1.727\times 10^{-16}$   &
  $1.028\times 10^{-5}$    & $1.993\times 10^{-16}$  \\
  50   &\hspace{0.8mm}  $6.031\times 10^{-6}$    & \hspace{0.8mm}   $1.168\times 10^{-16}$  &  \hspace{0.8mm} $8.225\times 10^{-6}$ & \hspace{0.8mm}
  $1.594\times 10^{-16}$   &
  $1.002\times 10^{-5}$    & $1.943\times 10^{-16}$  \\
  100  &\hspace{0.8mm}  $5.308\times 10^{-6}$    & \hspace{0.8mm}   $1.028\times 10^{-16}$  &  \hspace{0.8mm} $8.113\times 10^{-6}$ & \hspace{0.8mm}
  $1.572\times 10^{-16}$   &
  $9.988\times 10^{-6}$   & $1.935\times 10^{-16}$   \\
  300  &\hspace{0.8mm}  $4.285\times 10^{-6}$    & \hspace{0.8mm}   $8.304\times 10^{-17}$  &  \hspace{0.8mm} $8.032\times 10^{-6}$ & \hspace{0.8mm}
  $1.556\times 10^{-16}$   &
  $9.964\times 10^{-6}$   & $1.930\times 10^{-16}$   \\
  500  &\hspace{0.8mm}  $3.860\times 10^{-6}$    & \hspace{0.8mm}   $7.481\times 10^{-17}$  &  \hspace{0.8mm} $8.016\times 10^{-6}$ & $1.553\times 10^{-16}$ & $9.959\times 10^{-6}$  & $1.929\times 10^{-16} $     \\
\hline\hline

\multicolumn{7}{|c|}{ $95\%$ C.L. \hspace{5mm} $\sqrt{s}=380\hspace{0.8mm}GeV$ }\\
\hline\hline
\multicolumn{7}{|c|}{\hspace{2.5cm}  $\delta_{sys}=0\%$  \hspace{2.5cm}  $\delta_{sys}=5\%$  \hspace{2.5cm}  $\delta_{sys}=10\%$ }\\
\hline

\cline{1-7}  ${\cal L}\hspace{0.8mm}(fb^{-1})$  & \hspace{2mm} $|\mu_{\nu_\tau}(\mu_B)|$   & $\hspace{2mm} |d_{\nu_\tau}(e cm)|$
     & \hspace{2mm} $|\mu_{\nu_\tau}(\mu_B)|$   & $\hspace{2mm} |d_{\nu_\tau}(e cm)|$      & \hspace{2mm} $|\mu_{\nu_\tau}(\mu_B)|$
     & $\hspace{2mm} |d_{\nu_\tau}(e cm)|$         \\
\hline
  10   &\hspace{0.8mm}  $8.417\times 10^{-6}$    & \hspace{0.8mm}   $1.631\times 10^{-16}$  &  \hspace{0.8mm} $9.415\times 10^{-6}$  & \hspace{0.8mm}
  $1.824\times 10^{-16}$   &
  $1.084\times 10^{-5}$    & $2.101\times 10^{-16}$  \\
  50   &\hspace{0.8mm}  $6.418\times 10^{-6}$    & \hspace{0.8mm}   $1.243\times 10^{-16}$  &  \hspace{0.8mm} $8.700\times 10^{-6}$  & \hspace{0.8mm}
  $1.685\times 10^{-16}$   &
  $1.057\times 10^{-5}$    & $2.048\times 10^{-16}$  \\
  100  &\hspace{0.8mm}  $5.664\times 10^{-6}$    & \hspace{0.8mm}   $1.097\times 10^{-16}$  &  \hspace{0.8mm} $8.583\times 10^{-6}$  & \hspace{0.8mm}
  $1.663\times 10^{-16}$   &
  $1.053\times 10^{-5}$   & $2.040\times 10^{-16}  -$   \\
  300  &\hspace{0.8mm}  $4.593\times 10^{-6}$    & \hspace{0.8mm}   $8.901\times 10^{-17}$  &  \hspace{0.8mm} $8.499\times 10^{-6}$  & \hspace{0.8mm}
  $1.647\times 10^{-16}$   &
  $1.051\times 10^{-5}$   & $2.035\times 10^{-16}$   \\
  500  &\hspace{0.8mm}  $4.147\times 10^{-6}$    & \hspace{0.8mm}   $8.036\times 10^{-17}$  &  \hspace{0.8mm} $8.482\times 10^{-6}$  & $1.643\times 10^{-16}$  & $1.050\times 10^{-5}$  & $2.034\times 10^{-16}$     \\
\hline\hline
\end{tabular}
\end{center}
\end{table}

\begin{table}[!ht]
\caption{Sensitivity estimates on the $\mu_{\nu_\tau}$ magnetic moment and $d_{\nu_\tau}$ electric dipole moment via the process
$e^+ e^- \to \nu_\tau \bar\nu_\tau\gamma$ for $\sqrt{s}=1500\hspace{0.8mm}GeV$ and $P_{e^-}=P_{e^+}=0\%$.}
\begin{center}
\begin{tabular}{|c|cc|cc|cc|}
\hline\hline
\multicolumn{7}{|c|}{ $90\%$ C.L. \hspace{5mm} $\sqrt{s}=1500\hspace{0.8mm}GeV$ }\\
\hline\hline
\multicolumn{7}{|c|}{\hspace{2.5cm}  $\delta_{sys}=0\%$  \hspace{2.5cm}  $\delta_{sys}=5\%$  \hspace{2.5cm}  $\delta_{sys}=10\%$ }\\
\hline

\cline{1-7}  ${\cal L}\hspace{0.8mm}(fb^{-1})$  & \hspace{2mm} $|\mu_{\nu_\tau}(\mu_B)|$   & $\hspace{2mm} |d_{\nu_\tau}(e cm)|$
     & \hspace{2mm} $|\mu_{\nu_\tau}(\mu_B)|$   & $\hspace{2mm} |d_{\nu_\tau}(e cm)|$      & \hspace{2mm} $|\mu_{\nu_\tau}(\mu_B)|$
     & $\hspace{2mm} |d_{\nu_\tau}(e cm)|$         \\
\hline
  10   &\hspace{0.8mm}  $1.538\times 10^{-6}$    & \hspace{0.8mm}   $2.980\times 10^{-17}$  &  \hspace{0.8mm} $1.630\times 10^{-6}$ & \hspace{0.8mm}
  $3.160\times 10^{-17}$   &
  $1.826\times 10^{-6}$   & $3.539\times 10^{-17}$  \\
  100   &\hspace{0.8mm}  $9.145\times 10^{-7}$    & \hspace{0.8mm}   $1.772\times 10^{-17}$  &  \hspace{0.8mm} $1.268\times 10^{-6}$ & \hspace{0.8mm}
  $2.458\times 10^{-17}$   &
  $1.643\times 10^{-6}$   & $3.184\times 10^{-17}$  \\
  500  &\hspace{0.8mm}  $6.225\times 10^{-7}$    & \hspace{0.8mm}   $1.206\times 10^{-17}$  &  \hspace{0.8mm} $1.209\times 10^{-6}$ & \hspace{0.8mm}
  $2.343\times 10^{-17}$   &
  $1.622\times 10^{-6}$   & $3.143\times 10^{-17}$   \\
  1000  &\hspace{0.8mm}  $5.258\times 10^{-7}$    & \hspace{0.8mm}   $1.018\times 10^{-17}$  &  \hspace{0.8mm} $1.201\times 10^{-6}$ & \hspace{0.8mm}
  $2.327\times 10^{-17}$   &
  $1.619\times 10^{-6}$    & $3.138\times 10^{-17}$   \\
  1500  &\hspace{0.8mm}  $4.760\times 10^{-7}$    & \hspace{0.8mm}   $9.225\times 10^{-18}$  &  \hspace{0.8mm} $1.198\times 10^{-6}$ & $2.321\times 10^{-17}$ & $1.618\times 10^{-6}$  & $3.136\times 10^{-17} $     \\
\hline\hline

\multicolumn{7}{|c|}{ $95\%$ C.L. \hspace{5mm} $\sqrt{s}=1500\hspace{0.8mm}GeV$ }\\
\hline\hline
\multicolumn{7}{|c|}{\hspace{2.5cm}  $\delta_{sys}=0\%$  \hspace{2.5cm}  $\delta_{sys}=5\%$  \hspace{2.5cm}  $\delta_{sys}=10\%$ }\\
\hline

\cline{1-7}  ${\cal L}\hspace{0.8mm}(fb^{-1})$  & \hspace{2mm} $|\mu_{\nu_\tau}(\mu_B)|$   & $\hspace{2mm} |d_{\nu_\tau}(e cm)|$
     & \hspace{2mm} $|\mu_{\nu_\tau}(\mu_B)|$   & $\hspace{2mm} |d_{\nu_\tau}(e cm)|$      & \hspace{2mm} $|\mu_{\nu_\tau}(\mu_B)|$
     & $\hspace{2mm} |d_{\nu_\tau}(e cm)|$         \\
\hline
  10   &\hspace{0.8mm}  $1.656\times 10^{-6}$    & \hspace{0.8mm}   $3.210\times 10^{-17}$  &  \hspace{0.8mm} $1.754\times 10^{-6}$  & \hspace{0.8mm}
  $3.400\times 10^{-16}$   &
  $1.960\times 10^{-6}$   & $3.799\times 10^{-17}$  \\
  100   &\hspace{0.8mm}  $9.921\times 10^{-7}$    & \hspace{0.8mm}   $1.922\times 10^{-17}$  &  \hspace{0.8mm} $1.370\times 10^{-6}$  & \hspace{0.8mm}
  $2.656\times 10^{-17}$   &
  $1.767\times 10^{-6}$   & $3.425\times 10^{-17}$  \\
  500  &\hspace{0.8mm}  $6.772\times 10^{-7}$    & \hspace{0.8mm}   $1.312\times 10^{-17}$  &  \hspace{0.8mm} $1.307\times 10^{-6}$  & \hspace{0.8mm}
  $2.533\times 10^{-17}$   &
  $1.745\times 10^{-6}$   & $3.382\times 10^{-17}$   \\
  1000  &\hspace{0.8mm}  $5.724\times 10^{-7}$    & \hspace{0.8mm}   $1.109\times 10^{-17}$  &  \hspace{0.8mm} $1.298\times 10^{-6}$  & \hspace{0.8mm}
  $2.516\times 10^{-17}$   &
  $1.742\times 10^{-6}$    & $3.377\times 10^{-17}$   \\
  1500  &\hspace{0.8mm}  $5.185\times 10^{-7}$    & \hspace{0.8mm}   $1.004\times 10^{-17}$  &  \hspace{0.8mm} $1.295\times 10^{-6}$  & $2.510\times 10^{-17}$  & $1.741\times 10^{-6}$  & $3.375\times 10^{-17} $     \\
\hline\hline
\end{tabular}
\end{center}
\end{table}

\begin{table}[!ht]
\caption{Sensitivity estimates on the $\mu_{\nu_\tau}$ magnetic moment and $d_{\nu_\tau}$ electric dipole moment via the process
$e^+ e^- \to \nu_\tau \bar\nu_\tau\gamma$ for $\sqrt{s}=3000\hspace{0.8mm}GeV$ and $P_{e^-}=P_{e^+}=0\%$.}
\begin{center}
\begin{tabular}{|c|cc|cc|cc|}
\hline\hline
\multicolumn{7}{|c|}{ $90\%$ C.L. \hspace{5mm} $\sqrt{s}=3000\hspace{0.8mm}GeV$ }\\
\hline\hline
\multicolumn{7}{|c|}{\hspace{2.5cm}  $\delta_{sys}=0\%$  \hspace{2.5cm}  $\delta_{sys}=5\%$  \hspace{2.5cm}  $\delta_{sys}=10\%$ }\\
\hline

\cline{1-7}  ${\cal L}\hspace{0.8mm}(fb^{-1})$  & \hspace{2mm} $|\mu_{\nu_\tau}(\mu_B)|$   & $\hspace{2mm} |d_{\nu_\tau}(e cm)|$
     & \hspace{2mm} $|\mu_{\nu_\tau}(\mu_B)|$   & $\hspace{2mm} |d_{\nu_\tau}(e cm)|$      & \hspace{2mm} $|\mu_{\nu_\tau}(\mu_B)|$
     & $\hspace{2mm} |d_{\nu_\tau}(e cm)|$         \\
\hline
  100   &\hspace{0.8mm}  $4.834\times 10^{-7}$    & \hspace{0.8mm}   $9.368\times 10^{-18}$  &  \hspace{0.8mm} $5.593\times 10^{-7}$ & \hspace{0.8mm}
  $1.083\times 10^{-17}$   &
  $6.844\times 10^{-7}$   & $1.326\times 10^{-17}$  \\
  500  &\hspace{0.8mm}  $3.272\times 10^{-7}$    & \hspace{0.8mm}   $6.341\times 10^{-18}$  &  \hspace{0.8mm} $4.885\times 10^{-7}$ & \hspace{0.8mm}
  $1.081\times 10^{-17}$   &
  $6.532\times 10^{-7}$   & $1.265\times 10^{-17}$   \\
  1000  &\hspace{0.8mm}  $2.759\times 10^{-7}$    & \hspace{0.8mm}   $5.348\times 10^{-18}$  &  \hspace{0.8mm} $4.769\times 10^{-7}$ & \hspace{0.8mm}
  $9.242\times 10^{-18}$   &
  $6.489\times 10^{-7}$    & $1.257\times 10^{-17}$   \\
  2000  &\hspace{0.8mm}  $2.325\times 10^{-7}$    & \hspace{0.8mm}   $4.506\times 10^{-18}$  &  \hspace{0.8mm} $4.707\times 10^{-7}$ & $9.122\times 10^{-18}$ & $6.468\times 10^{-7}$  & $1.253\times 10^{-17} $     \\
  3000  &\hspace{0.8mm}  $2.103\times 10^{-7}$    & \hspace{0.8mm}   $4.076\times 10^{-18}$  &  \hspace{0.8mm} $4.686\times 10^{-7}$ & $9.081\times 10^{-18}$ & $6.460\times 10^{-7}$  & $1.251\times 10^{-17} $     \\

\hline\hline

\multicolumn{7}{|c|}{ $95\%$ C.L. \hspace{5mm} $\sqrt{s}=3000\hspace{0.8mm}GeV$ }\\
\hline\hline
\multicolumn{7}{|c|}{\hspace{2.5cm}  $\delta_{sys}=0\%$  \hspace{2.5cm}  $\delta_{sys}=5\%$  \hspace{2.5cm}  $\delta_{sys}=10\%$ }\\
\hline

\cline{1-7}  ${\cal L}\hspace{0.8mm}(fb^{-1})$  & \hspace{2mm} $|\mu_{\nu_\tau}(\mu_B)|$   & $\hspace{2mm} |d_{\nu_\tau}(e cm)|$
     & \hspace{2mm} $|\mu_{\nu_\tau}(\mu_B)|$   & $\hspace{2mm} |d_{\nu_\tau}(e cm)|$      & \hspace{2mm} $|\mu_{\nu_\tau}(\mu_B)|$
     & $\hspace{2mm} |d_{\nu_\tau}(e cm)|$         \\
\hline
  100   &\hspace{0.8mm}  $5.254\times 10^{-7}$    & \hspace{0.8mm}   $1.018\times 10^{-17}$  &  \hspace{0.8mm} $6.071\times 10^{-7}$  & \hspace{0.8mm}
  $1.176\times 10^{-17}$   &
  $7.414\times 10^{-7}$   & $1.436\times 10^{-17}$  \\
  500  &\hspace{0.8mm}  $3.563\times 10^{-7}$    & \hspace{0.8mm}   $6.905\times 10^{-18}$  &  \hspace{0.8mm} $5.309\times 10^{-7}$  & \hspace{0.8mm}
  $1.028\times 10^{-17}$   &
  $7.080\times 10^{-7}$   & $1.371\times 10^{-17}$   \\
  1000  &\hspace{0.8mm}  $3.007\times 10^{-7}$    & \hspace{0.8mm}   $5.826\times 10^{-18}$  &  \hspace{0.8mm} $5.183\times 10^{-7}$  & \hspace{0.8mm}
  $1.004\times 10^{-17}$   &
  $7.034\times 10^{-7}$    & $1.363\times 10^{-17}$   \\
  2000  &\hspace{0.8mm}  $2.534\times 10^{-7}$    & \hspace{0.8mm}   $4.912\times 10^{-18}$  &  \hspace{0.8mm} $5.116\times 10^{-7}$  & $9.915\times 10^{-18}$  & $7.010\times 10^{-7}$  & $1.358\times 10^{-17} $     \\
  3000  &\hspace{0.8mm}  $2.293\times 10^{-7}$    & \hspace{0.8mm}   $4.443\times 10^{-18}$  &  \hspace{0.8mm} $5.093\times 10^{-7}$  & $9.870\times 10^{-18}$  & $7.002\times 10^{-7}$  & $1.357\times 10^{-17} $     \\
\hline\hline
\end{tabular}
\end{center}
\end{table}

\subsection{Total cross section of the process $e^+ e^- \to \nu_\tau\bar\nu_\tau\gamma$ beyond the SM with polarized electron-positron beam}

Another option for sensitivty study of the total production of the channel $e^+ e^- \to \nu_\tau\bar\nu_\tau\gamma$, in addition
the dipole moments of the tau-neutrino, is the electron-positron beam polarization facility at the CLIC. The possibility of using
polarized electron and positron beams can constitute a strong advantage in searching for new physics \cite{Moortgat}. Furthermore,
the electron-positron beam polarization may lead to a reduction of the measurement uncertainties, either by increasing the signal
cross section, therefore reducing the statistical uncertainty, or by suppressing important backgrounds. In summary, one another
option at the CLIC is to polarize the incoming beams, which could maximize the physics potential, both in the performance of
precision tests and in revealing the properties of the new physics BSM.

The general formula for the total cross section for an arbitrary degree of longitudinal $e^-$ and $e^+$ beams polarization is give by \cite{Moortgat}

\begin{eqnarray}
\sigma(P_{e^-},P_{e^+})=&&\frac{1}{4}[(1+P_{e^-})(1+P_{e^+})\sigma_{++}+(1-P_{e^-})(1-P_{e^+})\sigma_{--}\nonumber\\
&&+(1+P_{e^-})(1-P_{e^+})\sigma_{+-}+(1-P_{e^-})(1+P_{e^+})\sigma_{-+}],
\end{eqnarray}

\noindent where $P_{e^-} (P_{e^+})$ is the polarization degree of the electron (positron) beam, while $\sigma_{-+}$
stands for the cross section for completely left-handed polarized $e^-$ beam $P_{e^-}=-1$ and completely right-handed
polarized $e^+$ beam $P_{e^+}=1$, and other cross sections $\sigma_{--}$, $\sigma_{++}$ and $\sigma_{+-}$ are defined
analogously.

For our sensitivity study, we assuming for definiteness an electron-positron beam polarization $(P_{e^-}, P_{e^+})=(-80\%, 60\%)$
in the estimated range of the expected CLIC operation setup. Besides the polarized beams we consider the isolation cuts given for
Eq. (8).

The numerical fit functions for the total cross sections of the process $e^+ e^- \to \nu_\tau\bar\nu_\tau\gamma$, following the
form of Eq.(9) with polarized electron-positron beam, and in terms of the independent form factors $F_2 (F_3)$ are given by:\\

\newpage

$\bullet$ For $\sqrt{s}=380\hspace{0.8mm} GeV$.

\begin{eqnarray}
\sigma(F_2)&=&\bigl[(3.97 \times 10^{11}) F^4_2 + (3.16 \times 10^{4} ) F^2_2 + 0.072\bigr](pb),   \nonumber \\
\sigma(F_3)&=&\bigl[(3.97 \times 10^{11}) F^4_3 + (3.16 \times 10^{4} ) F^2_3 + 0.072\bigr](pb).
\end{eqnarray}

$\bullet$ For $\sqrt{s}=1.5\hspace{0.8mm} TeV$.

\begin{eqnarray}
\sigma(F_2)&=&\bigl[(4.93 \times 10^{13}) F^4_2 + (7.23 \times 10^{5} ) F^2_2 + 0.023\bigr](pb),   \nonumber \\
\sigma(F_3)&=&\bigl[(4.93 \times 10^{13}) F^4_3 + (7.23 \times 10^{5} ) F^2_3 + 0.023\bigr](pb).
\end{eqnarray}

$\bullet$ For $\sqrt{s}=3\hspace{0.8mm} TeV$.

\begin{eqnarray}
\sigma(F_2)&=&\bigl[(2.23 \times 10^{14}) F^4_2 + (1.43 \times 10^{6} ) F^2_2 + 0.006\bigr](pb),   \nonumber \\
\sigma(F_3)&=&\bigl[(2.23 \times 10^{14}) F^4_3 + (1.43 \times 10^{6} ) F^2_3 + 0.006\bigr](pb).
\end{eqnarray}

\noindent In  Eqs. (15)-(17), the coefficients of $F_2 (F_3)$ given the anomalous contribution,
while the independent terms of $F_2 (F_3)$ correspond to the cross section at $F_2=F_3=0$ and represents the SM cross section.

\subsection{Sensitivity estimates on the $\mu_{\nu_\tau}$ and $d_{\nu_\tau}$  with polarized electron-positron beam}

The $e^+ e^- \to \nu_\tau\bar\nu_\tau\gamma$ production cross section, as a function of $F_2 (F_3)$ projected for the CLIC with
polarized electron-positron beam $(P_{e^-}, P_{e^+})=(-80\%, 60\%)$ and for the center-of-mass energies $\sqrt{s}=380, 1500,
3000\hspace{0.8mm}GeV$, they are shown in Figs. 7 and 8. From the direct comparison of Figs. 7 and 8, with their corresponding
for the unpolarized case Figs. 2 and 3, a significant gradual increase in the total production cross sections of 0.6, 60 and
$100\hspace{0.8mm}pb$ is clearly shown. In addition, the cross section increases with the increase of $F_2 (F_3)$, and decreases
as $F_2 (F_3)$ decreases. The SM result for the production cross section of the reaction $e^+ e^- \to \nu_\tau\bar\nu_\tau\gamma$
is obtained in the limit when $F_2 (F_3)=0$. In this case, the terms that depend on $F_2 (F_3)$ in Eqs. (15)-(17) are zero and
Eqs. (15)-(17) are reduced to the result for the SM.

Taking $(P_{e^-}, P_{e^+})=(-80\%, 60\%)$, $\sqrt{s}=380, 1500, 3000\hspace{0.8mm}GeV$ and ${\cal L}=10, 100, 500, 1500, 3000
\hspace{0.8mm}fb^{-1}$, the contours for estimate the sensitivity of $F_2$ and $F_3$ in the $F_2-F_3$ plane through the reaction
$e^+ e^- \to \nu_\tau\bar\nu_\tau\gamma$ are evaluated and shown in Figs. 9-11. Fig. 11 illustrates the better sensitivity for
$F_2$ and $F_3$ with $\sqrt{s}=3000\hspace{0.8mm}GeV$, ${\cal L}=10, 500, 3000 \hspace{0.8mm}fb^{-1}$ and $(P_{e^-}, P_{e^+})=(-80\%, 60\%)$.

Our results are given in Tables IV-VI, in which the sensitivity estimates on the $\mu_{\nu_\tau}$
and $d_{\nu_\tau}$ via the process $e^+ e^- \to \nu_\tau \bar\nu_\tau\gamma$ are shown for $P_{e^-}=-80\%$
and $P_{e^+}=60\%$, $\sqrt{s}=380, 1500, 3000\hspace{0.8mm}GeV$, ${\cal L}=10, 50, 100, 300, 500, 1000, 1500, 2000, 3000
\hspace{0.8mm}fb^{-1}$, $\delta_{sys}=0, 5, 10\hspace{1mm}\%$ at $90\%$ C.L. and $95\%$ C.L. The effect of the polarized
incoming $e^-$ and $e^+$ beams shows that the sensitivity on the $\mu_{\nu_\tau}$ and $d_{\nu_\tau}$
is enhanced by a $5\%$ at $P(-80\%;  60\%)$ polarization configuration, with respect to the unpolarized case
(see Tables I-III). Our most relevant results are: $|\mu_{\nu_\tau}(\mu_B)|=2.002\times 10^{-7}$ and $|d_{\nu_\tau}(e cm)|=4.039\times 10^{-18}$
at $90\%\hspace{0.8mm}C.L$.

\begin{table}[!ht]
\caption{Sensitivity estimates on the $\mu_{\nu_\tau}$ magnetic moment and $d_{\nu_\tau}$ electric dipole moment via the process
$e^+ e^- \to \nu_\tau \bar\nu_\tau\gamma$ for $\sqrt{s}=380\hspace{0.8mm}GeV$, $P_{e^-}=-80\%$ and $P_{e^-}=60\%$.}
\begin{center}
\begin{tabular}{|c|cc|cc|cc|}
\hline\hline
\multicolumn{7}{|c|}{ $90\%$ C.L. \hspace{5mm} $\sqrt{s}=380\hspace{0.8mm}GeV$ }\\
\hline\hline
\multicolumn{7}{|c|}{\hspace{2.5cm}  $\delta_{sys}=0\%$  \hspace{2.5cm}  $\delta_{sys}=5\%$  \hspace{2.5cm}  $\delta_{sys}=10\%$ }\\
\hline

\cline{1-7}  ${\cal L}\hspace{0.8mm}(fb^{-1})$  & \hspace{2mm} $|\mu_{\nu_\tau}(\mu_B)|$   & $\hspace{2mm} |d_{\nu_\tau}(e cm)|$
     & \hspace{2mm} $|\mu_{\nu_\tau}(\mu_B)|$   & $\hspace{2mm} |d_{\nu_\tau}(e cm)|$      & \hspace{2mm} $|\mu_{\nu_\tau}(\mu_B)|$
     & $\hspace{2mm} |d_{\nu_\tau}(e cm)|$         \\
\hline
  10   &\hspace{0.8mm}  $7.563\times 10^{-6}$    & \hspace{0.8mm}   $1.465\times 10^{-16}$  &  \hspace{0.8mm} $9.467\times 10^{-6}$ & \hspace{0.8mm}
  $1.592\times 10^{-16}$   &
  $1.059\times 10^{-5}$   & $1.735\times 10^{-16}$  \\
  50  &\hspace{0.8mm}  $5.684\times 10^{-6}$    & \hspace{0.8mm}   $1.101\times 10^{-16}$  &  \hspace{0.8mm} $8.485\times 10^{-6}$ & \hspace{0.8mm}
  $1.425\times 10^{-16}$   &
  $1.043\times 10^{-5}$   & $1.644\times 10^{-16}$   \\
  100  &\hspace{0.8mm}  $4.980\times 10^{-6}$    & \hspace{0.8mm}   $9.651\times 10^{-17}$  &  \hspace{0.8mm} $8.414\times 10^{-6}$ & \hspace{0.8mm}
  $1.394\times 10^{-16}$   &
  $1.041\times 10^{-5}$    & $1.394\times 10^{-16}$   \\
  300  &\hspace{0.8mm}  $3.993\times 10^{-6}$    & \hspace{0.8mm}   $7.737\times 10^{-17}$  &  \hspace{0.8mm} $8.365\times 10^{-6}$ & $1.372\times 10^{-16}$ & $1.039\times 10^{-5}$  & $1.621\times 10^{-16}$     \\
  500  &\hspace{0.8mm}  $3.585\times 10^{-6}$    & \hspace{0.8mm}   $6.948\times 10^{-17}$  &  \hspace{0.8mm} $8.355\times 10^{-6}$ & $1.367\times 10^{-16}$ & $1.038\times 10^{-5}$  & $1.619\times 10^{-16}$     \\

\hline\hline

\multicolumn{7}{|c|}{ $95\%$ C.L. \hspace{5mm} $\sqrt{s}=380\hspace{0.8mm}GeV$ }\\
\hline\hline
\multicolumn{7}{|c|}{\hspace{2.5cm}  $\delta_{sys}=0\%$  \hspace{2.5cm}  $\delta_{sys}=5\%$  \hspace{2.5cm}  $\delta_{sys}=10\%$ }\\
\hline

\cline{1-7}  ${\cal L}\hspace{0.8mm}(fb^{-1})$  & \hspace{2mm} $|\mu_{\nu_\tau}(\mu_B)|$   & $\hspace{2mm} |d_{\nu_\tau}(e cm)|$
     & \hspace{2mm} $|\mu_{\nu_\tau}(\mu_B)|$   & $\hspace{2mm} |d_{\nu_\tau}(e cm)|$      & \hspace{2mm} $|\mu_{\nu_\tau}(\mu_B)|$
     & $\hspace{2mm} |d_{\nu_\tau}(e cm)|$         \\
\hline
  10   &\hspace{0.8mm}  $8.018\times 10^{-6}$    & \hspace{0.8mm}   $1.553\times 10^{-16}$  &  \hspace{0.8mm} $9.467\times 10^{-6}$  & \hspace{0.8mm}
  $1.834\times 10^{-16}$   &
  $1.117\times 10^{-5}$   & $2.163\times 10^{-16}$  \\
  50  &\hspace{0.8mm}  $6.061\times 10^{-6}$    & \hspace{0.8mm}   $1.174\times 10^{-16}$  &  \hspace{0.8mm} $8.975\times 10^{-6}$  & \hspace{0.8mm}
  $1.739\times 10^{-16}$   &
  $1.099\times 10^{-5}$   & $2.130\times 10^{-16}$   \\
  100  &\hspace{0.8mm}  $5.326\times 10^{-6}$    & \hspace{0.8mm}   $1.032\times 10^{-16}$  &  \hspace{0.8mm} $8.902\times 10^{-6}$  & \hspace{0.8mm}
  $1.725\times 10^{-16}$   &
  $1.097\times 10^{-5}$    & $2.126\times 10^{-16}$   \\
  300  &\hspace{0.8mm}  $4.289\times 10^{-6}$    & \hspace{0.8mm}   $8.312\times 10^{-17}$  &  \hspace{0.8mm} $8.851\times 10^{-6}$  & $1.715\times 10^{-16}$  & $1.096\times 10^{-5}$  & $2.123\times 10^{-16} $     \\
  500  &\hspace{0.8mm}  $3.860\times 10^{-6}$    & \hspace{0.8mm}   $7.479\times 10^{-17}$  &  \hspace{0.8mm} $8.841\times 10^{-6}$  & $1.713\times 10^{-16}$  & $1.095\times 10^{-5}$  & $2.122\times 10^{-16}$     \\
\hline\hline
\end{tabular}
\end{center}
\end{table}

\begin{table}[!ht]
\caption{Sensitivity estimates on the $\mu_{\nu_\tau}$ magnetic moment and $d_{\nu_\tau}$ electric dipole moment via the process $e^+ e^- \to \nu_\tau \bar\nu_\tau\gamma$ for $\sqrt{s}=1500\hspace{0.8mm}GeV$, $P_{e^-}=-80\%$ and $P_{e^+}=60\%$.}
\begin{center}
\begin{tabular}{|c|cc|cc|cc|}
\hline\hline
\multicolumn{7}{|c|}{ $90\%$ C.L. \hspace{5mm} $\sqrt{s}=1500\hspace{0.8mm}GeV$ }\\
\hline\hline
\multicolumn{7}{|c|}{\hspace{2.5cm}  $\delta_{sys}=0\%$  \hspace{2.5cm}  $\delta_{sys}=5\%$  \hspace{2.5cm}  $\delta_{sys}=10\%$ }\\
\hline

\cline{1-7}  ${\cal L}\hspace{0.8mm}(fb^{-1})$  & \hspace{2mm} $|\mu_{\nu_\tau}(\mu_B)|$   & $\hspace{2mm} |d_{\nu_\tau}(e cm)|$
     & \hspace{2mm} $|\mu_{\nu_\tau}(\mu_B)|$   & $\hspace{2mm} |d_{\nu_\tau}(e cm)|$      & \hspace{2mm} $|\mu_{\nu_\tau}(\mu_B)|$
     & $\hspace{2mm} |d_{\nu_\tau}(e cm)|$         \\
\hline
  10   &\hspace{0.8mm}  $1.503\times 10^{-6}$    & \hspace{0.8mm}   $2.912\times 10^{-17}$  &  \hspace{0.8mm} $1.654\times 10^{-6}$ & \hspace{0.8mm}
  $3.206\times 10^{-17}$   &
  $1.926\times 10^{-6}$   & $3.732\times 10^{-17}$  \\
  100   &\hspace{0.8mm}  $8.940\times 10^{-7}$    & \hspace{0.8mm}   $1.732\times 10^{-17}$  &  \hspace{0.8mm} $1.379\times 10^{-6}$ & \hspace{0.8mm}
  $2.673\times 10^{-17}$   &
  $1.805\times 10^{-6}$   & $3.498\times 10^{-17}$  \\
  500  &\hspace{0.8mm}  $6.085\times 10^{-7}$    & \hspace{0.8mm}   $1.179\times 10^{-17}$  &  \hspace{0.8mm} $1.341\times 10^{-6}$ & \hspace{0.8mm}
  $2.600\times 10^{-17}$   &
  $1.792\times 10^{-6}$   & $3.474\times 10^{-17}$   \\
  1000  &\hspace{0.8mm}  $5.140\times 10^{-7}$    & \hspace{0.8mm}   $9.960\times 10^{-18}$  &  \hspace{0.8mm} $1.336\times 10^{-6}$ & \hspace{0.8mm}
  $2.590\times 10^{-17}$   &
  $1.791\times 10^{-6}$    & $3.471\times 10^{-17}$   \\
  1500  &\hspace{0.8mm}  $4.654\times 10^{-7}$    & \hspace{0.8mm}   $9.018\times 10^{-18}$  &  \hspace{0.8mm} $1.335\times 10^{-6}$ & $2.587\times 10^{-17}$ & $1.790\times 10^{-6}$  & $3.469\times 10^{-17} $     \\
\hline\hline

\multicolumn{7}{|c|}{ $95\%$ C.L. \hspace{5mm} $\sqrt{s}=1500\hspace{0.8mm}GeV$ }\\
\hline\hline
\multicolumn{7}{|c|}{\hspace{2.5cm}  $\delta_{sys}=0\%$  \hspace{2.5cm}  $\delta_{sys}=5\%$  \hspace{2.5cm}  $\delta_{sys}=10\%$ }\\
\hline

\cline{1-7}  ${\cal L}\hspace{0.8mm}(fb^{-1})$  & \hspace{2mm} $|\mu_{\nu_\tau}(\mu_B)|$   & $\hspace{2mm} |d_{\nu_\tau}(e cm)|$
     & \hspace{2mm} $|\mu_{\nu_\tau}(\mu_B)|$   & $\hspace{2mm} |d_{\nu_\tau}(e cm)|$      & \hspace{2mm} $|\mu_{\nu_\tau}(\mu_B)|$
     & $\hspace{2mm} |d_{\nu_\tau}(e cm)|$         \\
\hline
  10   &\hspace{0.8mm}  $1.618\times 10^{-6}$    & \hspace{0.8mm}   $3.137\times 10^{-17}$  &  \hspace{0.8mm} $1.779\times 10^{-6}$  & \hspace{0.8mm}
  $3.447\times 10^{-17}$   &
  $2.064\times 10^{-6}$   & $4.000\times 10^{-17}$  \\
  100   &\hspace{0.8mm}  $9.698\times 10^{-7}$    & \hspace{0.8mm}   $1.879\times 10^{-17}$  &  \hspace{0.8mm} $1.488\times 10^{-6}$  & \hspace{0.8mm}
  $2.883\times 10^{-17}$   &
  $1.937\times 10^{-6}$   & $3.754\times 10^{-17}$  \\
  500  &\hspace{0.8mm}  $6.620\times 10^{-7}$    & \hspace{0.8mm}   $1.282\times 10^{-17}$  &  \hspace{0.8mm} $1.448\times 10^{-6}$  & \hspace{0.8mm}
  $2.886\times 10^{-17}$   &
  $1.924\times 10^{-6}$   & $3.728\times 10^{-17}$   \\
  1000  &\hspace{0.8mm}  $5.596\times 10^{-7}$    & \hspace{0.8mm}   $1.084\times 10^{-17}$  &  \hspace{0.8mm} $1.442\times 10^{-6}$  & \hspace{0.8mm}
  $2.795\times 10^{-17}$   &
  $1.923\times 10^{-6}$    & $3.725\times 10^{-17}$   \\
  1500  &\hspace{0.8mm}  $5.069\times 10^{-7}$    & \hspace{0.8mm}   $9.823\times 10^{-18}$  &  \hspace{0.8mm} $1.440\times 10^{-6}$  & $2.792\times 10^{-17}$  & $1.922\times 10^{-6}$  & $3.724\times 10^{-17} $     \\
\hline\hline
\end{tabular}
\end{center}
\end{table}

\begin{table}[!ht]
\caption{Sensitivity estimates on the $\mu_{\nu_\tau}$ magnetic moment and $d_{\nu_\tau}$ electric dipole moment via the process
$e^+ e^- \to \nu_\tau \bar\nu_\tau\gamma$ for $\sqrt{s}=3000\hspace{0.8mm}GeV$, $P_{e^-}= -80\%$ and $P_{e^+}=60\%$.}
\begin{center}
\begin{tabular}{|c|cc|cc|cc|}
\hline\hline
\multicolumn{7}{|c|}{ $90\%$ C.L. \hspace{5mm} $\sqrt{s}=3000\hspace{0.8mm}GeV$ }\\
\hline\hline
\multicolumn{7}{|c|}{\hspace{2.5cm}  $\delta_{sys}=0\%$  \hspace{2.5cm}  $\delta_{sys}=5\%$  \hspace{2.5cm}  $\delta_{sys}=10\%$ }\\
\hline

\cline{1-7}  ${\cal L}\hspace{0.8mm}(fb^{-1})$  & \hspace{2mm} $|\mu_{\nu_\tau}(\mu_B)|$   & $\hspace{2mm} |d_{\nu_\tau}(e cm)|$
     & \hspace{2mm} $|\mu_{\nu_\tau}(\mu_B)|$   & $\hspace{2mm} |d_{\nu_\tau}(e cm)|$      & \hspace{2mm} $|\mu_{\nu_\tau}(\mu_B)|$
     & $\hspace{2mm} |d_{\nu_\tau}(e cm)|$         \\
\hline
  100   &\hspace{0.8mm}  $4.609\times 10^{-7}$    & \hspace{0.8mm}   $9.276\times 10^{-18}$  &  \hspace{0.8mm} $5.735\times 10^{-7}$ & \hspace{0.8mm}
  $1.160\times 10^{-17}$   &
  $7.284\times 10^{-7}$   & $1.475\times 10^{-17}$  \\
  500  &\hspace{0.8mm}  $3.116\times 10^{-7}$    & \hspace{0.8mm}   $6.281\times 10^{-18}$  &  \hspace{0.8mm} $5.235\times 10^{-7}$ & \hspace{0.8mm}
  $1.064\times 10^{-17}$   &
  $7.085\times 10^{-7}$   & $1.437\times 10^{-17}$   \\
  1000  &\hspace{0.8mm}  $2.628\times 10^{-7}$    & \hspace{0.8mm}   $5.299\times 10^{-18}$  &  \hspace{0.8mm} $5.161\times 10^{-7}$ & \hspace{0.8mm}
  $1.049\times 10^{-17}$   &
  $7.059\times 10^{-7}$    & $1.432\times 10^{-17}$   \\
  2000  &\hspace{0.8mm}  $2.214\times 10^{-7}$    & \hspace{0.8mm}   $4.465\times 10^{-18}$  &  \hspace{0.8mm} $5.122\times 10^{-7}$ & $1.042\times 10^{-17}$ & $7.046\times 10^{-7}$  & $1.429\times 10^{-17} $     \\
  3000  &\hspace{0.8mm}  $2.002\times 10^{-7}$    & \hspace{0.8mm}   $4.039\times 10^{-18}$  &  \hspace{0.8mm} $5.109\times 10^{-7}$ & $1.039\times 10^{-17}$ & $7.041\times 10^{-7}$  & $1.428\times 10^{-17} $     \\

\hline\hline

\multicolumn{7}{|c|}{ $95\%$ C.L. \hspace{5mm} $\sqrt{s}=3000\hspace{0.8mm}GeV$ }\\
\hline\hline
\multicolumn{7}{|c|}{\hspace{2.5cm}  $\delta_{sys}=0\%$  \hspace{2.5cm}  $\delta_{sys}=5\%$  \hspace{2.5cm}  $\delta_{sys}=10\%$ }\\
\hline

\cline{1-7}  ${\cal L}\hspace{0.8mm}(fb^{-1})$  & \hspace{2mm} $|\mu_{\nu_\tau}(\mu_B)|$   & $\hspace{2mm} |d_{\nu_\tau}(e cm)|$
     & \hspace{2mm} $|\mu_{\nu_\tau}(\mu_B)|$   & $\hspace{2mm} |d_{\nu_\tau}(e cm)|$      & \hspace{2mm} $|\mu_{\nu_\tau}(\mu_B)|$
     & $\hspace{2mm} |d_{\nu_\tau}(e cm)|$         \\
\hline
  100   &\hspace{0.8mm}  $5.010\times 10^{-7}$    & \hspace{0.8mm}   $1.007\times 10^{-17}$  &  \hspace{0.8mm} $6.223\times 10^{-7}$  & \hspace{0.8mm}
  $1.258\times 10^{-17}$   &
  $7.883\times 10^{-7}$   & $1.595\times 10^{-17}$  \\
  500  &\hspace{0.8mm}  $3.394\times 10^{-7}$    & \hspace{0.8mm}   $6.840\times 10^{-18}$  &  \hspace{0.8mm} $5.686\times 10^{-7}$  & \hspace{0.8mm}
  $1.154\times 10^{-17}$   &
  $7.671\times 10^{-7}$   & $1.554\times 10^{-17}$   \\
  1000  &\hspace{0.8mm}  $2.863\times 10^{-7}$    & \hspace{0.8mm}   $5.773\times 10^{-18}$  &  \hspace{0.8mm} $5.605\times 10^{-7}$  & \hspace{0.8mm}
  $1.139\times 10^{-17}$   &
  $7.642\times 10^{-7}$    & $1.549\times 10^{-17}$   \\
  2000  &\hspace{0.8mm}  $2.413\times 10^{-7}$    & \hspace{0.8mm}   $4.867\times 10^{-18}$  &  \hspace{0.8mm} $5.564\times 10^{-7}$  & $1.131\times 10^{-17}$  & $7.628\times 10^{-7}$  & $1.546\times 10^{-17} $     \\
  3000  &\hspace{0.8mm}  $2.183\times 10^{-7}$    & \hspace{0.8mm}   $4.403\times 10^{-18}$  &  \hspace{0.8mm} $5.549\times 10^{-7}$  & $1.128\times 10^{-17}$  & $7.623\times 10^{-7}$  & $1.545\times 10^{-17} $     \\
\hline\hline
\end{tabular}
\end{center}
\end{table}

\section{Conclusions}

In this paper, we have sensitivity estimates on the total cross section and on the dipole moments $\mu_{\nu_\tau}$ and $d_{\nu_\tau}$
through the process $e^+ e^- \to \nu_\tau \bar\nu_\tau\gamma$ at the future CLIC. Furthermore, the process is analyzed for two scenarios
motivated by the strong advantage in searching for new physics BSM: a) unpolarized electron-positron beam $(P_{e^-}, P_{e^+})=(0, 0)$ and
b) polarized  electron-positron beam $(P_{e^-}, P_{e^+})=(-80\%, 60\%)$. In the first scenario, the unpolarized cross section has the
value of $\approx 200 (200)\hspace{0.8mm} pb$ (see Figs. 2 and 3) depending on the anomalous coupling type $F_2$ $(F_3)$. In the second
scenario, which is motivated by the possibility to enhance or suppress different physical processes, the polarized cross section gets a
value of $\approx 300 (300) pb$ (see Figs. 7 and 8) depending on the anomalous coupling type $F_2$ $(F_3)$. Comparing each scenario shows
that the cross section is enhanced for $100\hspace{0.8mm}pb$ for the case of polarized electron-positron beam. The option of upgrading the
incoming electron and the positron beam to be polarized has the power to enhance the potential of the machine. In addition to these, the
results for the sensitivity contours in the $F_2-F_3$ plane for the unpolarized and polarized case are presented (see Figs. 4-6 and 9-11).

Figs. 2-11 and Tables I-VI highlight that sensitivity estimates for the total cross section, the form factors $F_2 (F_3)$, as well as for
the anomalous $\mu_{\nu_\tau}$ and $d_{\nu_\tau}$ at CLIC for high center-of-mass energies and
high luminosities, they reach a better sensitivity to that of L3 \cite{L3}, CERN-WA-066 \cite{A.M.Cooper} and E872 (DONUT) \cite{DONUT}, as
well as of others experimental and theoretical results (see Table I of Ref. \cite{Gutierrez13}). The most optimistic scenario about the
sensitivity in the anomalous dipole moments of the $\tau$-neutrino (see Tables III and VI), yields the following results:
$|\mu_{\nu_\tau}(\mu_B)|=2.103\times 10^{-7}$ and $|d_{\nu_\tau}(e cm)|=4.076\times10^{-18}$ with $(P_{e^-}, P_{e^+})=(0\%, 0\%)$. In addition,
we also obtain the results $|\mu_{\nu_\tau}(\mu_B)|=2.002\times 10^{-7}$ and $|d_{\nu_\tau}(e cm)|=4.039\times 10^{-18}$ with
$(P_{e^-}, P_{e^+})=(-80\%, 60\%)$. Our results show the potential and the feasibility of the process $e^+ e^- \to \nu_\tau \bar\nu_\tau\gamma$
at the CLIC.

In conclusion, the process itself is very useful to sensitivity probing on the dipole moments of the tau-neutrino and illustrates
the complementarity between CLIC and other $e^+e^-$ and pp colliders for probing extensions of the SM. Furthermore, we hope that
this work will motivate further studies of the $e^+ e^- \to \nu_\tau \bar\nu_\tau\gamma$ process, using in particular polarized
electro-positron beams.


\vspace*{0.5cm}

\begin{center}
{\bf Acknowledgments}
\end{center}%

A. G. R. and M. A. H. R. acknowledges support from SNI and PROFOCIE (M\'exico).

\newpage

\pagebreak

\begin{figure}[t]
\centerline{\scalebox{1}{\includegraphics{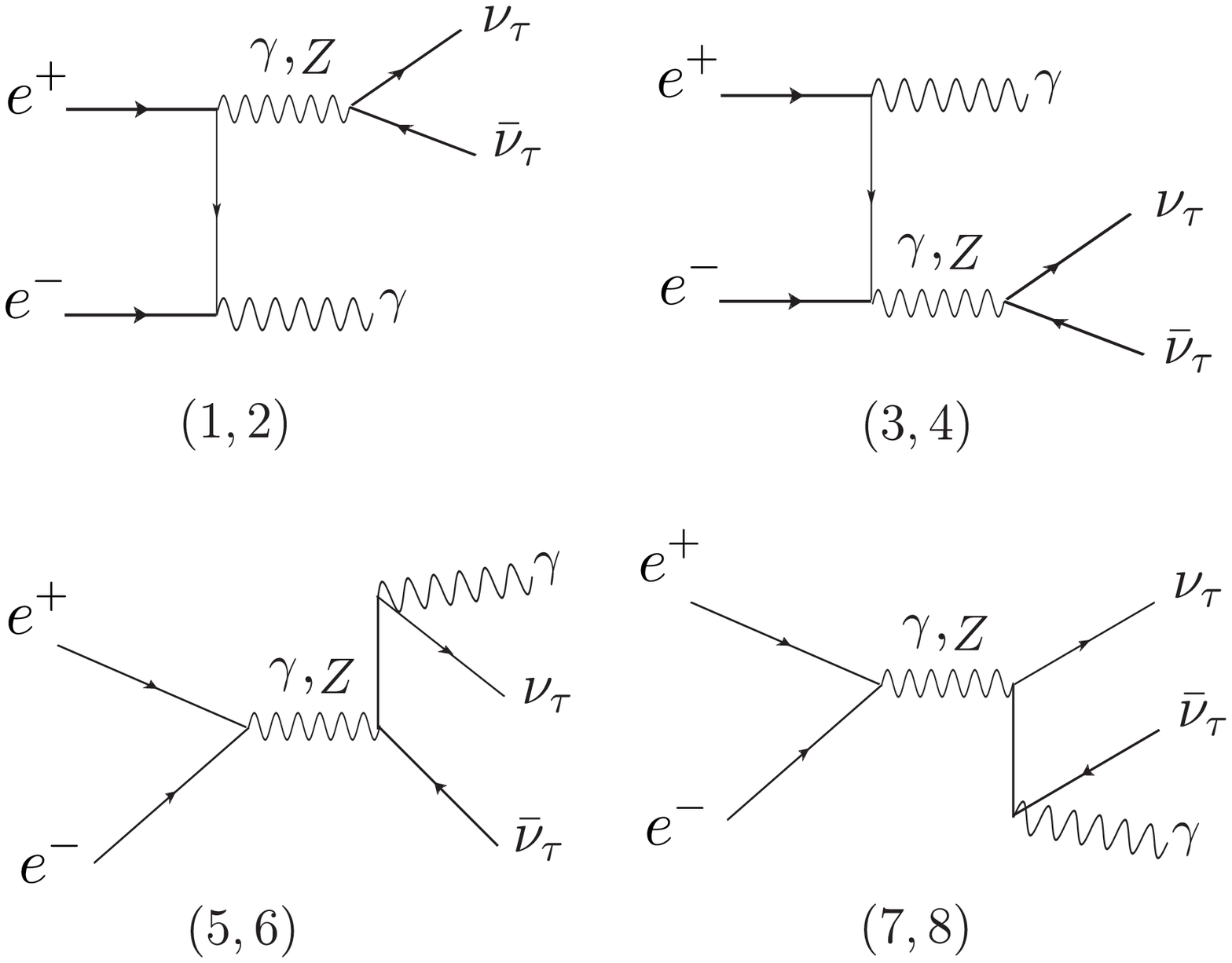}}}
\caption{ \label{fig:gamma2} The Feynman diagrams for the process
$e^+ e^-\rightarrow (\gamma, Z) \to \nu_\tau \bar \nu_\tau \gamma$.}
\label{Fig.2}
\end{figure}

\begin{figure}[t]
\centerline{\scalebox{1.6}{\includegraphics{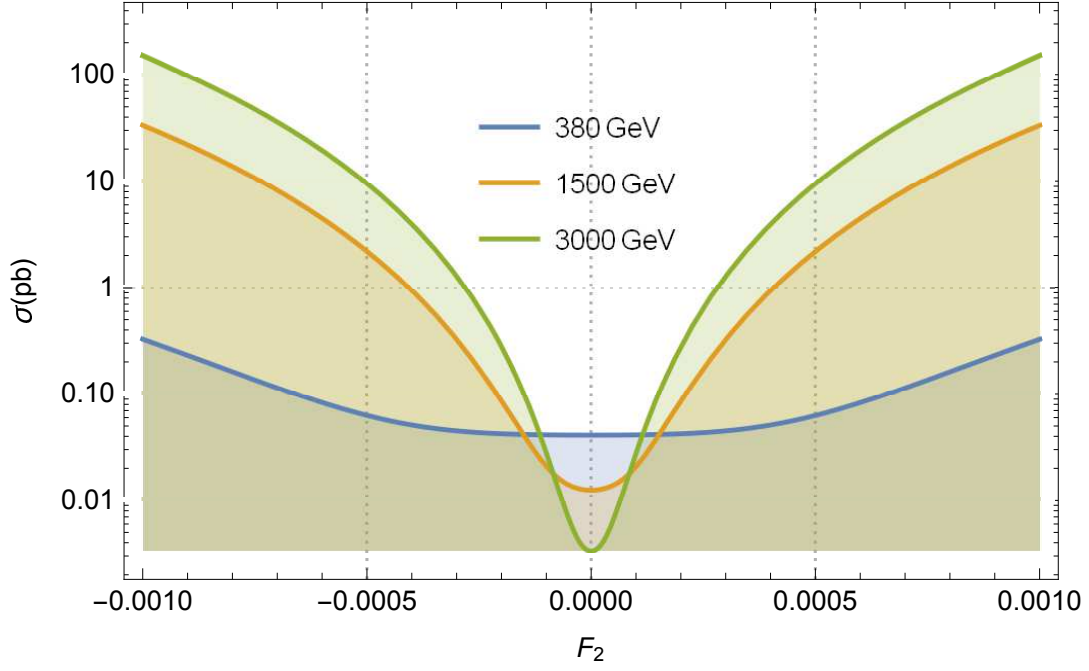}}}
\caption{ \label{fig:gamma1} The total cross sections of the process
$e^+e^-\rightarrow \nu_\tau \bar \nu_\tau \gamma$ as a function of $F_2$
for center-of-mass energies of $\sqrt{s}=380, 1500, 3000$\hspace{0.8mm}$GeV$.}
\label{Fig.3}
\end{figure}

\begin{figure}[t]
\centerline{\scalebox{1.6}{\includegraphics{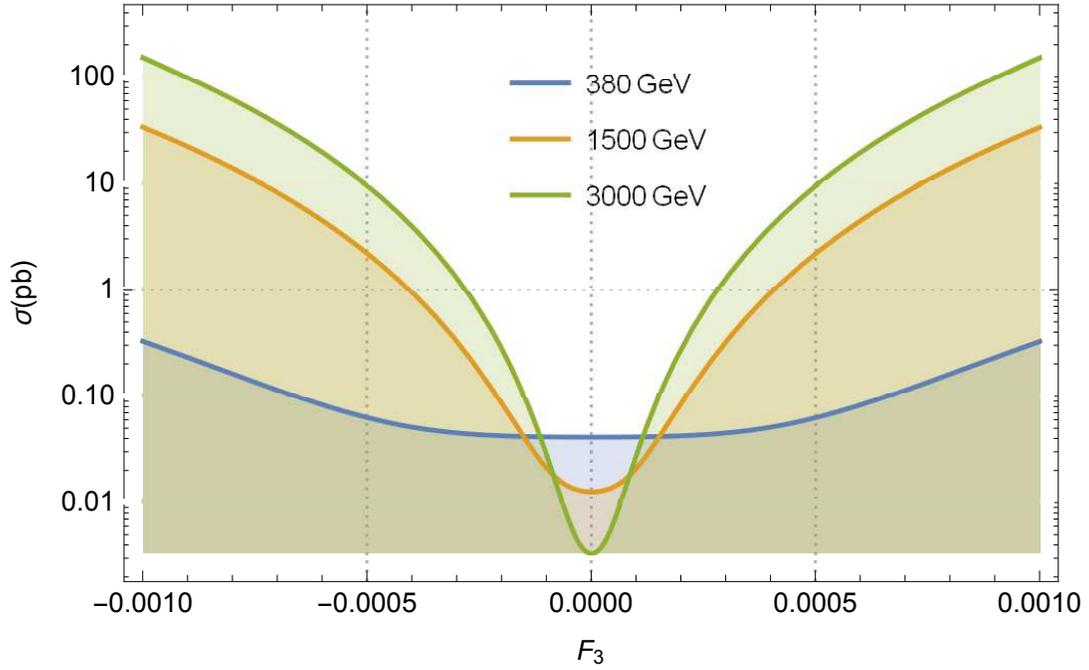}}}
\caption{ \label{fig:gamma2} Same as in Fig. 3, but for $F_3$.}
\label{Fig.4}
\end{figure}

\begin{figure}[t]
\centerline{\scalebox{1.2}{\includegraphics{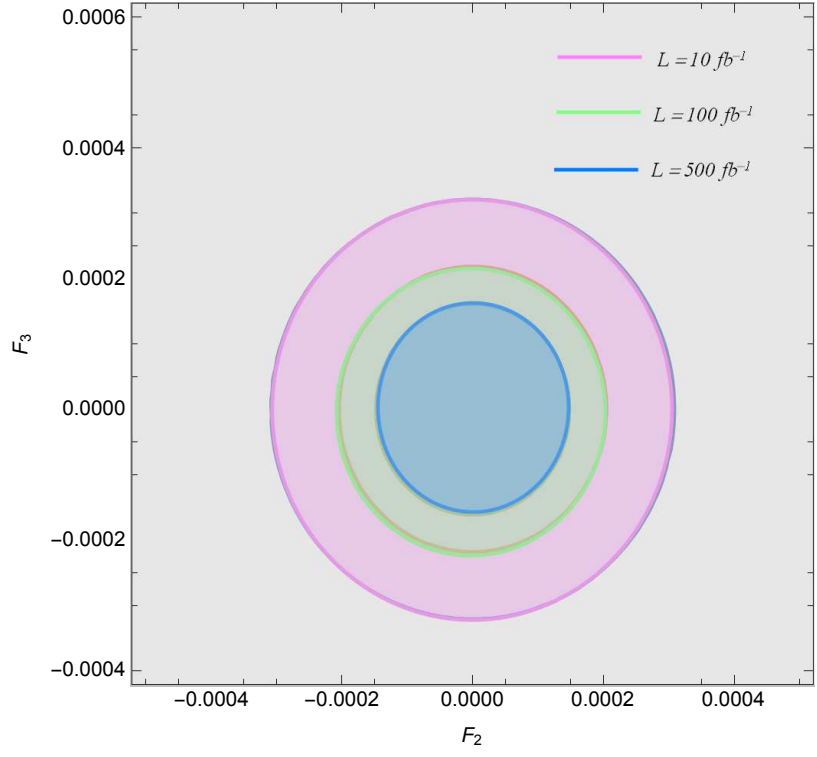}}}
\caption{ \label{fig:gamma3} Sensitivity contours at the $95\% \hspace{1mm}C.L.$ in the
$F_3-F_2$ plane for the process $e^+e^-\rightarrow \nu_\tau \bar \nu_\tau \gamma$
with the $\delta _{sys}=0\%$ and for center-of-mass energy of $\sqrt{s}=380\hspace{0.8mm}GeV$.}
\label{Fig.5}
\end{figure}

\begin{figure}[t]
\centerline{\scalebox{1.2}{\includegraphics{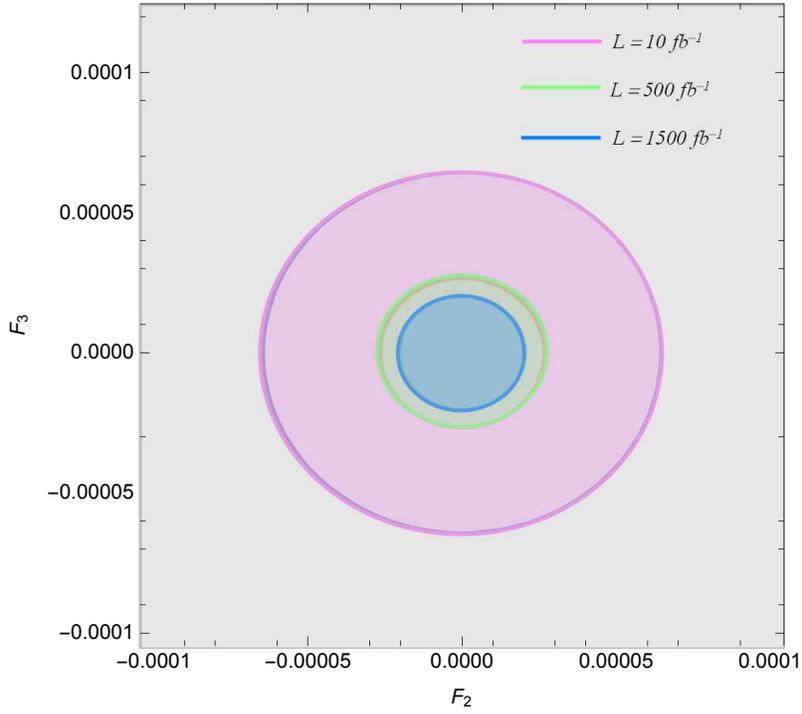}}}
\caption{ \label{fig:gamma15} Same as in Fig. 4, but for $\sqrt{s}=1500\hspace{0.8mm}GeV$.}
\label{Fig.6}
\end{figure}

\begin{figure}[t]
\centerline{\scalebox{1.2}{\includegraphics{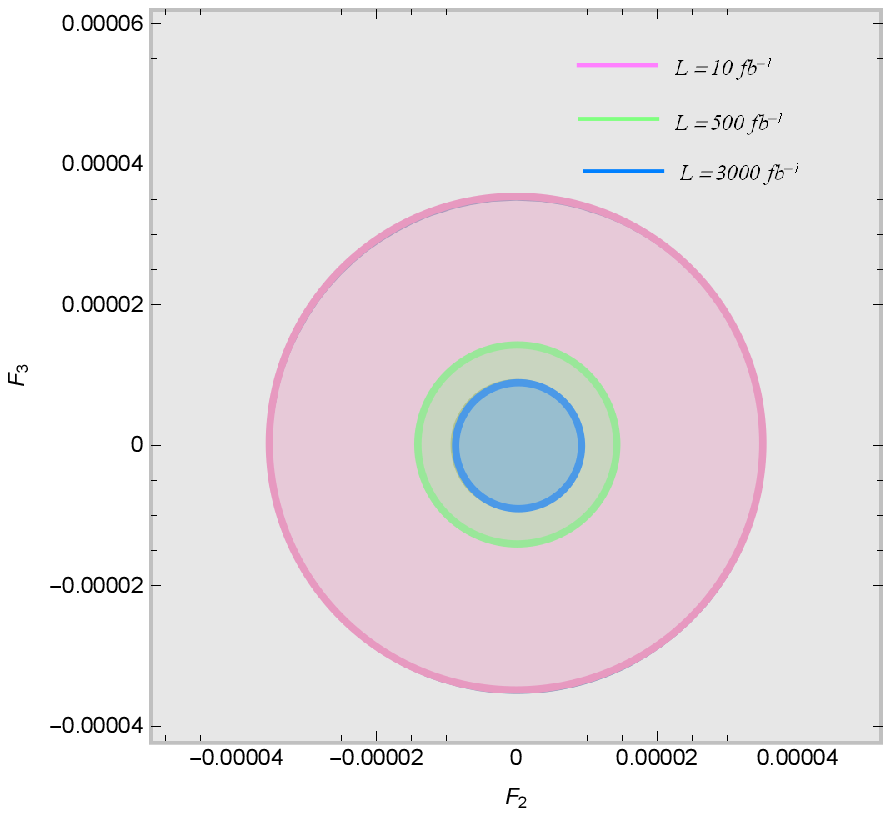}}}
\caption{ \label{fig:gamma15} Same as in Fig. 4, but for $\sqrt{s}=3000\hspace{0.8mm}GeV$.}
\label{Fig.6}
\end{figure}

\begin{figure}[t]
\centerline{\scalebox{1.6}{\includegraphics{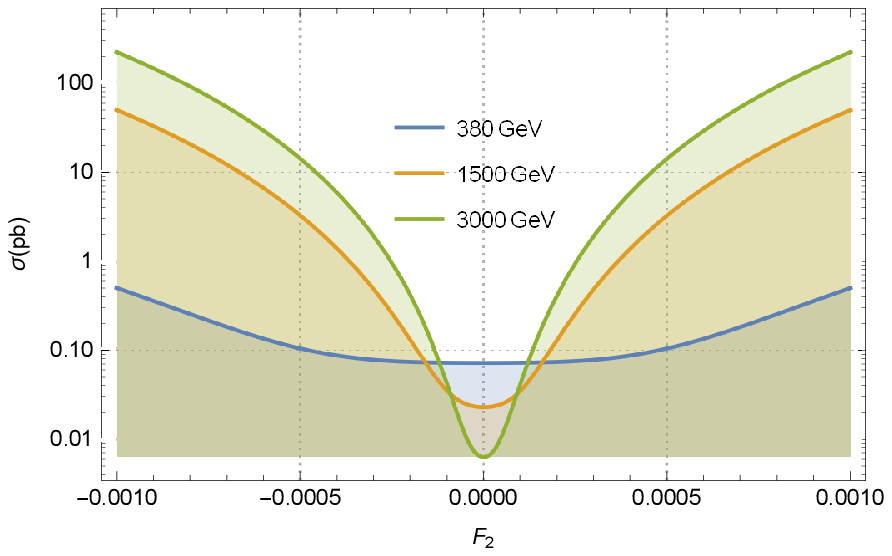}}}
\caption{ \label{fig:gamma15} Same as in Fig. 2, but for $P_{e^-}=-80\%$ and $P_{e^+}=60\%$.}
\label{Fig.6}
\end{figure}

\begin{figure}[t]
\centerline{\scalebox{1.6}{\includegraphics{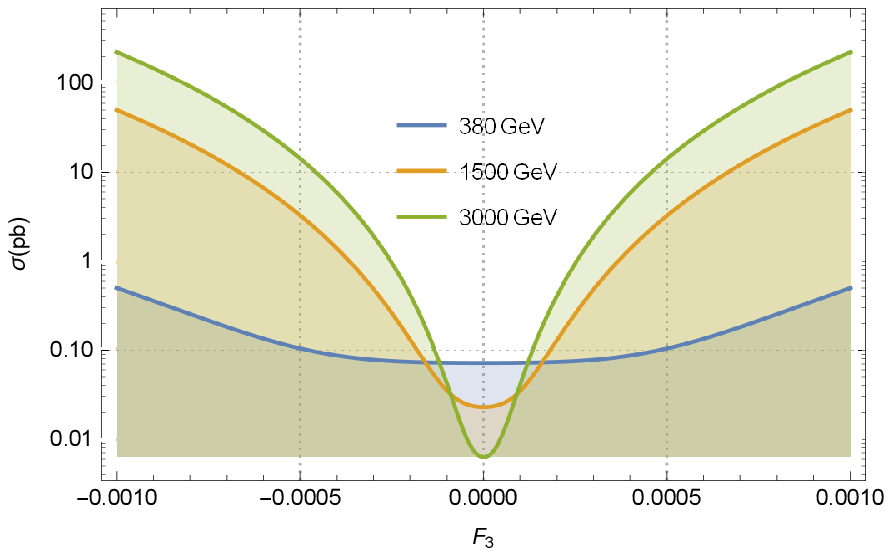}}}
\caption{ \label{fig:gamma15} Same as in Fig. 3, but for $P_{e^-}=-80\%$ and $P_{e^+}=60\%$.}
\label{Fig.6}
\end{figure}

\begin{figure}[t]
\centerline{\scalebox{1.2}{\includegraphics{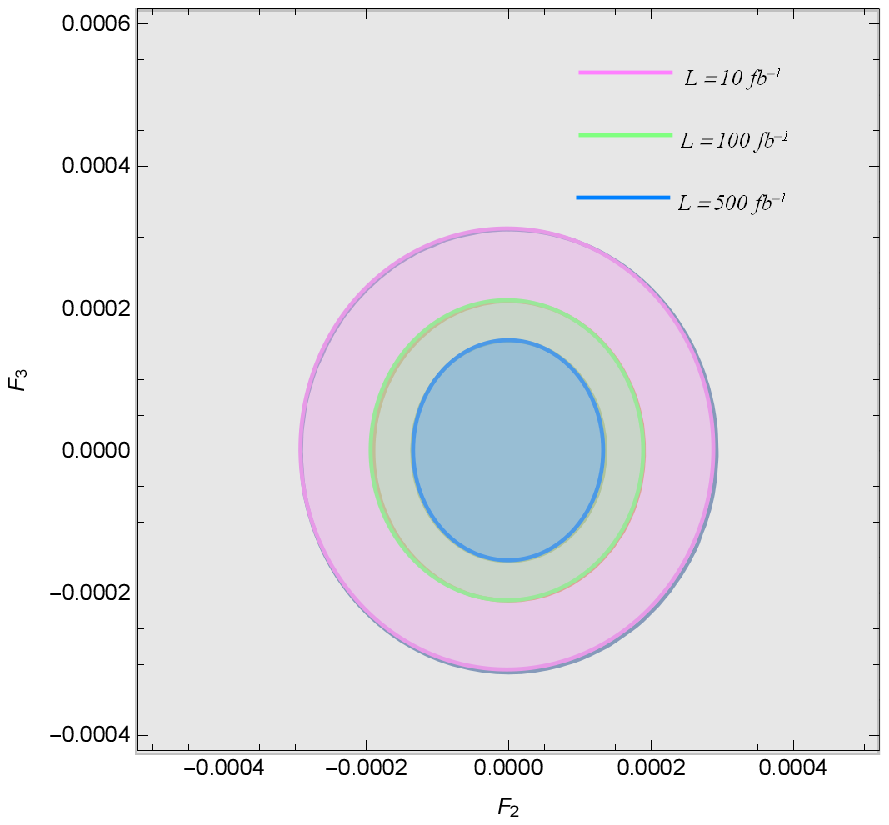}}}
\caption{ \label{fig:gamma15} Same as in Fig. 4, but for $P_{e^-}=-80\%$ and $P_{e^+}=60\%$.}
\label{Fig.6}
\end{figure}

\begin{figure}[t]
\centerline{\scalebox{1.2}{\includegraphics{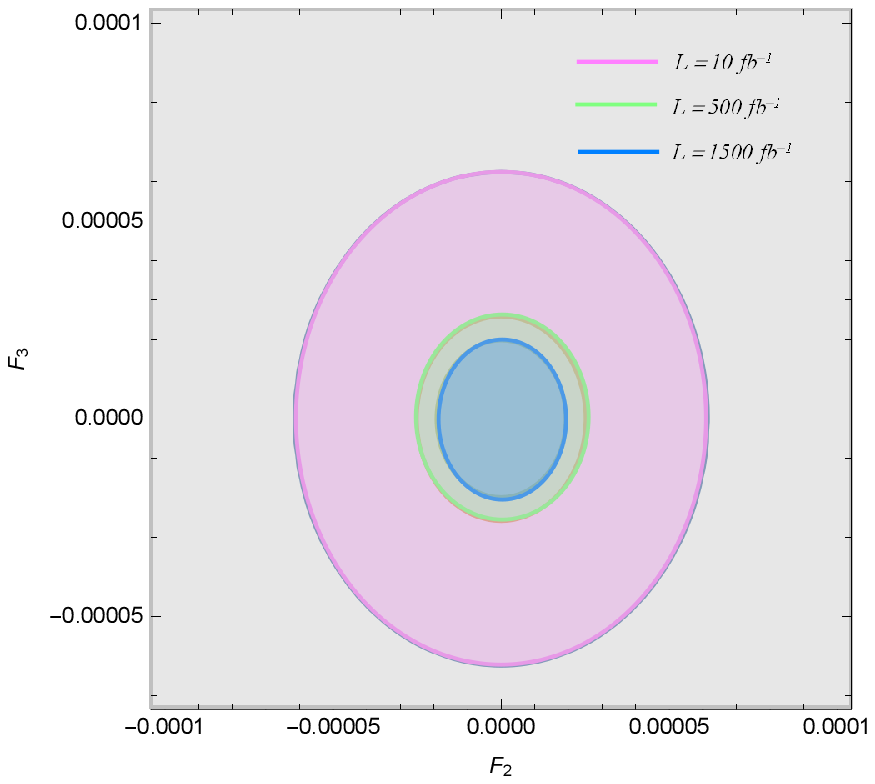}}}
\caption{ \label{fig:gamma15} Same as in Fig. 5, but for $P_{e^-}=-80\%$ and $P_{e^+}=60\%$.}
\label{Fig.6}
\end{figure}

\begin{figure}[t]
\centerline{\scalebox{1.2}{\includegraphics{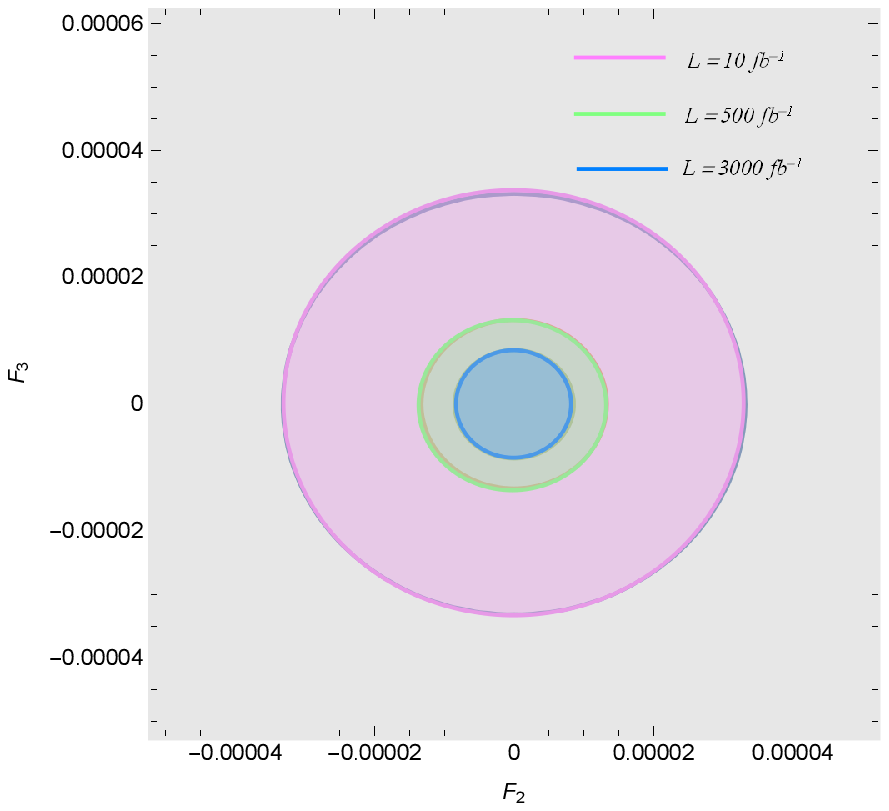}}}
\caption{ \label{fig:gamma15} Same as in Fig. 6, but for $P_{e^-}=-80\%$ and $P_{e^+}=60\%$.}
\label{Fig.6}
\end{figure}

\end{document}